\documentclass[a4paper,10pt]{article}
\usepackage[dvips]{graphicx}
\begin{document}
\title{Phase Transition in a Three-States Reaction-Diffusion System}
\author{F H Jafarpour\footnote{Corresponding author's e-mail:farhad@ipm.ir} \, and B Ghavami \\ \\
{\small Bu-Ali Sina University, Physics Department, Hamadan, Iran} }
\maketitle
\begin{abstract}
A one-dimensional reaction-diffusion model consisting of two species
of particles and vacancies on a ring is introduced. The number of
particles in one species is conserved while in the other species it
can fluctuate because of creation and annihilation of particles. It
has been shown that the model undergoes a continuous phase
transition from a phase where the currents of different species of
particles are equal to another phase in which they are different.
The total density of particles and also their currents in each phase
are calculated exactly.
\end{abstract}
\maketitle %
Recently one-dimensional reaction-diffusion systems have received
much attention because they show a variety of interesting critical
phenomena such as out-of-equilibrium phase transitions
\cite{sz,sch}. A simple system of this type, which has been studied
widely in related literatures, is the Asymmetric Simple Exclusion
Process (ASEP) \cite{dehp}. In this two-states model the particles
are injected from the left site of an open discreet lattice of
length $L$. They diffuse in the system and at the end of the lattice
are extracted from the system. It is known that depending on the
injection and the extraction rates the ASEP shows different boundary
induced phase transitions. Non-equilibrium phase transition may also
happen in the systems with non-conserving dynamics \cite{hsi,eklm}.
For instance in \cite{eklm} the authors investigate a three-states
model consists of two species of particles besides vacancies on a
lattice with ring geometry. The dynamics of this model consists of
diffusion, creation and annihilation of both species of the
particles. They have found that the phase diagram of the model
highly depends on the annihilation rate of the particles. By
changing the annihilation rate of the particles, the system
transfers from a maximal current phase to a fluid phase. The density
of the vacancies changes discontinuously from one phase to the other
phase. In present paper we introduce and study a reaction-diffusion
model on a discrete lattice of length $L$ with periodic boundary
condition. Besides the vacancies there are two different types of
particles in the system. Throughout this paper the vacancies and the
particles are denoted by E, A and B. The dynamics of the system is
not conserving. The particles of type A and type B hop to the left
and to the right respectively. The total number of particles of type
B is a conserved quantity and assumed to be equal to $M$. The
density of these particles is defined as $\rho_B=\frac{M}{L}$. In
contrast, the total number of particles of type A is not a conserved
quantity due to the creation and annihilation of them. Only the
nearest neighbors interactions are allowed and the model evolves
through the following processes
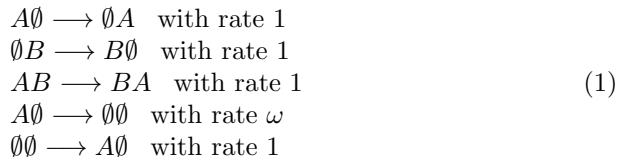
\begin{equation}
\label{Process}
\begin{array}{l}
A \emptyset \longrightarrow \emptyset A \;\;\; \mbox{with rate 1} \\ %%%%%%
\emptyset B \longrightarrow B \emptyset  \;\;\; \mbox{with rate 1} \\ %%%%%%
A B \longrightarrow B A \;\;\; \mbox{with rate 1}  \\ %%%%%%
A \emptyset \longrightarrow \emptyset \emptyset \;\;\; \mbox{with rate $\omega$} \\ %%%%%%
\emptyset \emptyset \longrightarrow  A \emptyset   \;\;\; \mbox{with rate 1} %%%%%
\end{array}
\end{equation}
As can be seen the parameter $\omega$ determines the annihilation
rate for the particles of type A which besides the number of the
particles of type B i.e. $\rho_B$ are the free parameters of the
model. One should note that the annihilation in our model only takes
place for one species of particles. Our main aim in the present work
is to study the phase diagram of the model in terms of $\omega$ and
the density of the B particles.\\
In our model if one starts with a lattice without any vacancies the
dynamics of the model prevents it from evolving into other
configurations consisting of vacancies. In this case the system
remains in its initial configuration and the steady state of the
system is trivial. In order to study the non-trivial case we
consider those configurations which have at least one vacancy. In
order to find the stationary probability distribution function of
the system we apply the Matrix Product Formalism (MPF) first
introduced in \cite{dehp} and then generalized in \cite{ks}.
According to this formalism the stationary probability for any
configuration of a system with periodic boundary condition is
proportional to the trace of product of non-commuting operators
which satisfy a quadratic algebra. In our model we have three
different states at each site of the lattice associated with the
presence of vacancies, the A particles and the B particles. We
assign three different operators $\bf E$, $\bf A$ and $\bf B$ to
each state. Now the unnormalized steady state probability of a
configuration $\mathcal{C}$ is given by
\begin{equation}
\label{SPDF} P(\mathcal{C})=\frac{1}{\mathcal{Z}_L}
Tr[\prod_{i=1}^{L}{\bf X}_i]
\end{equation}
in which ${\bf X}_i={\bf E}$ if the site $i$ is empty, ${\bf
X}_i={\bf A}$ if the site $i$ is occupied by a particle of type A
and ${\bf X}_i={\bf B}$ if it is occupied by a particle of type B.
The normalization factor $\mathcal{Z}_L$ in the denominator of
(\ref{SPDF}) is called the partition function of the system and is
given by the sum of unnormalized weights of all accessible
configurations. By applying the MPF one finds the following
quadratic algebra for our model
\begin{equation}
\label{ALG1}
\begin{array}{l}
{\bf A} {\bf B}={\bf A}+{\bf B} \\
{\bf A} {\bf E}={\bf E}\\
{\bf E} {\bf B}={\bf E}\\
{\bf E}^2=\omega {\bf E}.
\end{array}
\end{equation}
Now by defining ${\bf E}=\omega \vert V \rangle \langle W \vert$ in
which $\langle W \vert V \rangle=1$ one can simply find
\begin{equation}
\label{ALG2}
\begin{array}{l}
{\bf A} {\bf B}={\bf A}+{\bf B} \\
{\bf A}\vert V \rangle= \vert V \rangle\\
\langle W \vert {\bf B}=\langle W \vert\\
{\bf E}=\omega \vert V \rangle \langle W \vert.
\end{array}
\end{equation}
The first three relations in (\ref{ALG2}) make a quadratic algebra
which is well known in the related literatures. It is the quadratic
algebra of the ASEP when the boundary rates are equal to
one\footnote{The operator {\bf A} and {\bf B} should be regarded as
the operators associated with the particles and vacancies
respectively.}. This algebra has an infinite dimensional
representation given by the following matrices and vectors
\begin{equation}
{\bf A}=\left(
    \begin{array}{ccccc}
      1 & 1 & 0 & 0 & \cdots \\
      0 & 1 & 1 & 0 &  \\
      0 & 0 & 1 & 1 &  \\
      0 & 0 & 0 & 1 &  \\
      \vdots &   &   &   & \ddots \\
    \end{array}
  \right), {\bf B}={\bf A}^T, \vert V \rangle=\left(
                                    \begin{array}{c}
                                      1 \\
                                      0 \\
                                      0 \\
                                      0 \\
                                      \vdots \\
                                    \end{array}
                                  \right),
   \langle W \vert =\vert V \rangle ^T
\end{equation}
in which $T$ stands for transpose. In order to find the phase
structure of the system one can calculate the generating function of
the partition function of the system and study its singularities. In
what follows we first calculate the grandcanonical partition
function of the system $\mathcal{Z}_L(\xi)$ by introducing a
fugacity $\xi$ for particles of type B. We then fix the fugacity of
the B particles using the following relation
\begin{equation}
\label{Fugacity Relation} \rho_B=\lim_{L\rightarrow \infty}
\frac{\xi}{L}\frac{\partial \ln \mathcal{Z}_L(\xi)}{\partial \xi}.
\end{equation}
The grandcanonical partition function of the system
$\mathcal{Z}(\xi)$ can now be calculated from (\ref{SPDF}) and is
given by
\begin{equation}
\label{GCPF}
\mathcal{Z}_{L}(\xi)=\sum_{\mathcal{C}}Tr[\prod_{i=1}^{L}{\bf
X}_i]=Tr[({\bf A}+\xi {\bf B}+{\bf E})^L-({\bf A}+\xi {\bf B})^L].
\end{equation}
One should note that the operator ${\bf B}$ in (\ref{SPDF}) is
replaced with the operator $\xi {\bf B}$ in which $\xi$ should be
fixed using (\ref{Fugacity Relation}). As we mentioned above the
stationary state of the system without vacancies is a trivial one,
therefore in (\ref{GCPF}) we have considered those configurations
with at least one vacancy. The generating function for
$\mathcal{Z}_L(\xi)$ can now be calculated using (\ref{ALG2}). Using
the same procedure introduced in \cite{srs} and after some
straightforward algebra one finds
\begin{equation}
\label{GF} \mathcal{G}(\xi,\lambda)=\sum_{L=0}^{\infty}\lambda^{L-1}
\mathcal{Z}_{L}(\xi)=\frac{\frac{d}{d\lambda} \omega
U(\xi,\lambda)}{1-\omega U(\xi,\lambda)}
\end{equation}
in which
\begin{equation}
U(\xi,\lambda)=\sum_{L=0}^{\infty}\lambda^{L+1}\langle W \vert ({\bf
A}+\xi {\bf B})^L \vert V \rangle.
\end{equation}
The convergence radius of the formal series (\ref{GF}) which is the
absolute value of its nearest singularity to the origin can be
written as
\begin{equation}
R(\xi)=\lim_{L\rightarrow
\infty}{\mathcal{Z}_{L}(\xi)}^\frac{-1}{L}.
\end{equation}
This is also the inverse of the largest eigenvalue of
$\mathcal{Z}_{L}(\xi)$. In the large $L$ limit using (\ref{Fugacity
Relation}) this results in the following relation
\begin{equation}
\label{Fugacity Relation 2} \rho_B=\xi\frac{\partial}{\partial
\xi}\ln\frac{1}{R(\xi)}.
\end{equation}
Therefore one should only find the singularities of (\ref{GF}) and
decide in which region of the phase diagram which singularity is the
smallest one. In order to find the singularities of (\ref{GF}) one
should first calculate $U(\xi,\lambda)$. This can easily be done by
noting that the matrix ${\bf A}+\xi {\bf B}$ defined as
\begin{equation}
{\bf A}+\xi {\bf B}=\left(
    \begin{array}{ccccc}
      1+\xi & 1 & 0 & 0 & \cdots \\
      \xi & 1+\xi & 1 & 0 &  \\
      0 & \xi & 1+\xi & 1 &  \\
      0 & 0 & \xi & 1+\xi &  \\
      \vdots &   &   &   & \ddots \\
    \end{array}
  \right)
\end{equation}
satisfy the following eigenvalue relation
\begin{equation}
\label{eigenvalue} ({\bf A}+\xi {\bf B}) \vert \theta \rangle =
(1+\xi+2\sqrt{\xi}Cos(\theta))\vert \theta \rangle
\end{equation}
for $-\pi \leq \theta \leq \pi$ in which we have defined
\begin{equation}
\vert \theta \rangle=\left(\begin{array}{c}
Sin(\theta) \\
\xi^{1/2}Sin(2\theta) \\
\xi Sin(3\theta)\\
\xi^{3/2}Sin(4\theta) \\
\xi^{2}Sin(5\theta) \\
\vdots \\
\end{array}
\right).
\end{equation}
Considering the fact that
$\int_{-\pi}^{\pi}\frac{d\theta}{\pi}Sin(\theta)Sin(n\theta)=\delta_{1,n}$
one can easily see that
\begin{equation}
\label{V} \vert V \rangle =
\int_{-\pi}^{\pi}\frac{d\theta}{\pi}Sin(\theta)\vert \theta \rangle.
\end{equation}
Now using (\ref{eigenvalue})-(\ref{V}) and the fact that $\langle W
\vert \theta \rangle =Sin(\theta)$ one can easily show that
$U(\xi,\lambda)$ is given by
\begin{equation}
\label{INTREP}
U(\xi,\lambda)=\int_{-\pi}^{\pi}\frac{d\theta}{\pi}\frac{\lambda
Sin(\theta)^2}{1-\lambda(1+\xi+2\sqrt{\xi}Cos(\theta))}
\end{equation}
which is valid for $\lambda < \frac{1}{(1+{\sqrt\xi})^2}$. The
integral (\ref{INTREP}) can easily be calculated using the Cauchy
residue theorem by noting that it has three poles in the complex
plane. Two of them are inside the contour which is a circle of unite
radius around the origin and one is outside it. After some
calculations one finds
\begin{equation}
U(\xi,\lambda)=\frac{1-\lambda-\lambda
\xi-\sqrt{(\lambda+\lambda\xi-1)^2-4\lambda^2\xi}}{2\lambda\xi}.
\end{equation}
Now that $U(\xi,\lambda)$ is calculated one can simply find the
singularities of (\ref{GF}). It turns out that (\ref{GF}) has two
different kinds of singularities: a simple root singularity
$R_1=\frac{\omega}{(1+\omega)(\omega+\xi)}$ which come from the
denominator of (\ref{GF}) and a square root singularity
$R_2=\frac{1}{(1+\sqrt{\xi})^2}$. Therefore the model has two
different phases. The relation between the density of B particles
and their fugacity in each phase should be obtained from
(\ref{Fugacity Relation 2}). In terms of the density of the B
particles $\rho_B$ we find $R_1=\frac{1-\rho_B}{1+\omega}$ and
$R_2=(1-\rho_B)^2$. Two different scenarios might happen: defining
$\omega_c=\frac{\rho_B}{1-\rho_B}$ we find that for $\omega >
\omega_c$ the nearest singularity to the origin is $R_1$ and for
$\omega < \omega_c$ it is $R_2$. The density of the vacancies can be
calculated quit similar to that of the B particles. It is given by
\begin{equation}
\label{Fugacity Relation 3}
\rho_E=\omega\frac{\partial}{\partial
\omega}\ln\frac{1}{R(\xi)}
\end{equation}
in which $R(\xi)$ is again the nearest singularity to the origin in
each phase. The density of the A particles is in turn
$\rho_A=1-\rho_B-\rho_E$. Let us now investigate the current of the
particles in each phase. Noting that the configurations without
vacancies are inaccessible, the particle current for each species is
obtained to be
\begin{eqnarray}
J_{\bf A}& =& \frac{Tr[(\xi {\bf A}{\bf B}+{\bf A}{\bf E})({\bf
A}+\xi {\bf B} + {\bf E})^{L-2}-(\xi {\bf A}{\bf B})({\bf A}+\xi
{\bf B})^{L-2}]}
{Tr[({\bf A}+\xi {\bf B}+{\bf E})^L-({\bf A}+\xi {\bf B})^L]}\\
J_{{\bf B}}& =& \frac{Tr[(\xi {\bf A}{\bf B}+\xi {\bf E}{\bf
B})({\bf A}+\xi {\bf B} + {\bf E})^{L-2}-(\xi {\bf A}{\bf B})({\bf
A}+\xi {\bf B})^{L-2}]}{Tr[({\bf A}+\xi {\bf B}+{\bf E})^L-({\bf
A}+\xi {\bf B})^L]}.
\end{eqnarray}
These relations can be simplified in the thermodynamic limit and one
finds
\begin{eqnarray}
J_{A}& =& R(\xi)(1+(\xi-1)\rho_A)\\
J_{B}& =& R(\xi)(\xi+(1-\xi)\rho_B).
\end{eqnarray}
As we mentioned above for $\omega < \omega_c$ the nearest
singularity to the origin is always $R_2$. The particle currents in
this case are equal and we find $J_{A}=J_{B}=\rho_B(1-\rho_B)$. In
contrast for $\omega > \omega_c$ the nearest singularity to the
origin is always $R_1$ and it turns out that the currents are not
equal. We find $J_A=\frac{\omega}{(1+\omega)^2}$ and
$J_B=\rho_B(1-\rho_B)$. In what follows we bring the summery of the
results concerning the phase structure of the system
$$
\mbox{for $\omega < \omega_c$}\;\;\; \left\{
\begin{array}{ll}
\rho_A=1-\rho_B& J_A=\rho_B(1-\rho_B) \\
\rho_B=\rho_B& J_B=\rho_B(1-\rho_B)\\
\rho_E=0&
\end{array}
\right.
$$
and
$$
\mbox{for $\omega >
\omega_c$}\;\;\; \left\{
\begin{array}{ll}
\rho_A=\frac{1}{1+\omega}&J_A=\frac{\omega}{(1+\omega)^2} \\
\rho_B=\rho_B& J_B=\rho_B(1-\rho_B)\\
\rho_E=\frac{\omega}{1+\omega}-\rho_B&
\end{array}
\right. .
$$
As can be seen for $\omega < \omega_c$ the density of vacancies is
equal to zero which means there are only A and B particles on the
lattice. Since the density of the B particles is fixed and equal to
$\rho_B$ the density of A particles should be $1-\rho_B$. In this
case, according to (\ref{Process}), both A and B particles have
simply ASEP dynamics and therefore their currents should be of the
form $\rho(1-\rho)$ and that $J_A=J_B$. This is in quite agreement
with our calculations for $J_A$ and $J_B$. On the other hand for
$\omega > \omega_c$ the density of vacancies on the lattice is no
longer zero. At the transition point $\omega_c$ the density of the
vacancies is zero but it increases linearly in this phase. In terms
of the density of the vacancies the phase transition is a continuous
transition. In this phase for $\omega < 1$ we always have $J_A >
J_B$ while for $\omega > 1$ we have $J_A > J_B$ for $\rho_B <
\frac{1}{1+\omega}$ and $J_A < J_B$
for $\frac{1}{1+\omega}< \rho_B < \frac{\omega}{1+\omega}$.\\
In this paper we studied a three-states model consists of A and B
particles besides the vacancies. The A particles are created and
annihilated which is controlled by $\omega$ while the B particles
only diffuse on the lattice and have a fixed density $\rho_B$. We
found that the system have two phases depending on $\rho_B$ and
$\omega$. The current of the B particles is always a constant
$\rho_B(1-\rho_B)$ throughout the phase diagram while for the A
particles it is given by different expressions in each phase. As a
generalization one could also consider a more general process
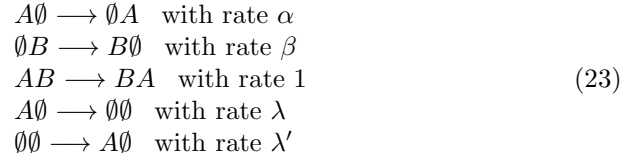
\begin{equation}
\begin{array}{l}
A \emptyset \longrightarrow \emptyset A \;\;\; \mbox{with rate $\alpha$} \\ %%%%%%
\emptyset B \longrightarrow B \emptyset  \;\;\; \mbox{with rate $\beta$} \\ %%%%%%
A B \longrightarrow B A \;\;\; \mbox{with rate 1}  \\ %%%%%%
A \emptyset \longrightarrow \emptyset \emptyset \;\;\; \mbox{with rate $\lambda$} \\ %%%%%%
\emptyset \emptyset \longrightarrow  A \emptyset   \;\;\; \mbox{with rate $\lambda '$} %%%%%
\end{array}
\end{equation}
with the quadratic algebra given by
\begin{equation}
\begin{array}{l}
{\bf A} {\bf B}={\bf A}+{\bf B} \\
\alpha \; {\bf A}\vert V \rangle= \vert V \rangle\\
\beta \; \langle W \vert {\bf B}=\langle W \vert\\
{\bf E}=\frac{\lambda}{\lambda ' \alpha} \vert V \rangle \langle W
\vert.
\end{array}
\end{equation}
and apply the same approach used in present paper to study its phase
diagram. This is under our investigations.


\begin{thebibliography}{1}

\bibitem{sz} B. Schmittmann and R. K. P. Zia, {\em Phase Transitions and Critical
Phenomena}, Vol 17, C. Domb and J. Lebowitz eds. (Academic, London,
1994)

\bibitem{sch} G. M. Sch\"utz {\em Phase Transitions and Critical Phenomena}, Vol 19,
C. Domb and J. Lebowitz eds. (Academic, London, 2001)

\bibitem{dehp} B. Derrida, M.R. Evans, V. Hakim and V. Pasquier {\it J. Phys. A: Math. Gen.} A {\bf 26} 1493 (1993)

\bibitem{hsi} H. Hinrichsen, S. Sandow and I. Peschel {\it J. Phys. A: Math. Gen.} A {\bf 29} 2643 (1996)

\bibitem{eklm} M. R. Evans, Y. Kafri, E. Levine, and D. Mukamel {\it J. Phys. A: Math. Gen.} A {\bf 35} L433 (2002)

\bibitem{ks} K. Krebs and S. Sandow {\it J. Phys. A: Math. Gen.} A {\bf 30} 3165 (1997)

\bibitem{srs} N. Rajewsky, T. Sasamoto and E. R. Speer {\it Physica A} {\bf 279} 123 (2000)

\end{thebibliography}
\end{document}